\documentclass[twocolumn,showpacs,preprintnumbers,superscriptaddress,aps,prx]{revtex4-2}

\usepackage{graphicx}
\usepackage{amsmath}
\usepackage{amssymb}
\usepackage{bm}
\usepackage{braket}
\usepackage{cprotect}
\usepackage{lmodern}
\usepackage{ulem}
\usepackage{listings,jvlisting}
\usepackage{physics}
\usepackage{xspace}
\usepackage{booktabs}
\usepackage{tikz}

\usepackage{dcolumn}

\usepackage{bm}
\newcommand{\correction}[1]{\textcolor{black}{#1}}

\begin{document}

\preprint{}

\title{Helical Magnetic State in the Vicinity of the Pressure-Induced Superconducting Phase in MnP}

\author{Sachith E. Dissanayake}
\altaffiliation[Present address: ]{Department of Mechanical Engineering, University of Rochester, Rochester, New York, 14617, USA}
\affiliation{Neutron Scattering Division, Oak Ridge National Laboratory, Oak Ridge, Tennessee 37831, USA}

\author{Masaaki Matsuda}
\altaffiliation[Corresponding author:] { matsudam@ornl.gov}
\affiliation{Neutron Scattering Division, Oak Ridge National Laboratory, Oak Ridge, Tennessee 37831, USA}

\author{Kazuyoshi Yoshimi}
\affiliation{Institute for Solid State Physics, University of Tokyo, Kashiwa, Chiba 277-8581, Japan}

\author{Shusuke Kasamatsu}
\affiliation{Faculty of Science, Yamagata University, 1-4-12 Kojirakawa, Yamagata-shi, Yamagata 990-8560, Japan}

\author{Feng Ye}
\affiliation{Neutron Scattering Division, Oak Ridge National Laboratory, Oak Ridge, Tennessee 37831, USA}

\author{Songxue Chi}
\affiliation{Neutron Scattering Division, Oak Ridge National Laboratory, Oak Ridge, Tennessee 37831, USA}

\author{William Steinhardt}
\affiliation{Department of Physics, Duke University, Durham, NC 27708, USA}

\author{Gilberto Fabbris}
\affiliation{Advanced Photon Source, Argonne National Laboratory, Argonne, Illinois 60439, USA}

\author{Sara Haravifard}
\affiliation{Department of Physics, Duke University, Durham, NC 27708, USA}
\affiliation{Department of Materials Sciences and Mechanical Engineering, Duke University, Durham, NC 27708, USA}

\author{Jinguang Cheng}
\affiliation{Beijing National Laboratory for Condensed Matter Physics and Institute of Physics, Chinese Academy of Sciences, Beijing 100190, China}

\author{Jiaqiang Yan}
\affiliation{Materials Science and Technology Division, Oak Ridge National Laboratory, Oak Ridge, Tennessee 37831, USA}

\author{Jun Gouchi}
\affiliation{Institute for Solid State Physics, University of Tokyo, Kashiwa, Chiba 277-8581, Japan}

\author{Yoshiya Uwatoko}
\affiliation{Institute for Solid State Physics, University of Tokyo, Kashiwa, Chiba 277-8581, Japan}

\date{\today}

\begin{abstract}
MnP is a metal that shows successive magnetic transitions from paramagnetic to ferromagnetic and helical magnetic phases at ambient pressure with decreasing temperature. With applied pressure, the magnetic transition temperatures decrease and superconductivity appears around 8 GPa where the magnetic order is fully suppressed and the quantum critical behavior is observed.
These results suggest that MnP is an unconventional superconductor in which magnetic fluctuations may be relevant to the superconducting pairing mechanism.
In order to elucidate the magnetic ground state adjacent to the superconducting phase first discovered in Mn-based materials, high-pressure neutron diffraction measurements have been performed in hydrostatic pressure up to \correction{7.5} GPa. The helical magnetic structure with the propagation vector along the $b$ axis, reported previously at 3.8 GPa, was found to be robust up to \correction{7.5} GPa.
First principles and classical Monte Carlo calculations have also been performed to understand how the pressure-driven magnetic phase transitions are coupled with change of the exchange interactions.
The calculations, which qualitatively reproduce the magnetic structures as a function of pressure, suggest that the exchange interactions change drastically with applied pressure and the further-neighbor interactions become more influential at high pressures.
Combining the experimental and theoretical results, we describe
the detail of exchange interactions in the vicinity of the superconducting phase which is critical to understand the pairing mechanism of the unconventional superconductivity in MnP.
\end{abstract}

\maketitle

\section{Introduction}
Unconventional superconductivity in high transition temperature superconductors, like cuprates \cite{Keimer2015}, iron-based \cite{Dai2012}, and heavy-fermion superconductors \cite{Pfleiderer2009} has been investigated intensively. Superconductivity appears in the quantum critical region, where the long-range magnetic order is suppressed and non-Fermi liquid behavior appears.
Thus magnetic fluctuations are potentially related to the superconducting pairing mechanism.

Pressure-induced superconductivity was reported in two materials with the orthorhombic MnP-type structure with space group $Pnma$ in the last decade. CrAs orders with a helical magnetic structure below 270 K at ambient pressure.
Pressure suppresses magnetic order, and a superconducting phase emerges between 0.3 and 2.1 GPa with a maximum ordering temperature of 2 K \cite{Wu2014,Kotegawa2014}.
Quantum critical behavior was reported around the pressure where the helical magnetic order is suppressed \cite{Wu2014,Kotegawa2014,Matsuda2018}. These results suggest that the magnetic fluctuations are related to the superconducting pairing mechanism in CrAs.

MnP was also found to be superconducting below $\sim$1 K around 8 GPa \cite{Cheng2015}.
There are only two known Mn-based superconductors, the other being KMn$_6$Bi$_5$ with a maximum transition temperature of $\sim$9 K~\cite{Liu2022}. This rarity is due to the strong magnetic pair-breaking effect of the sizable Mn magnetic moment.
As in CrAs, this system exhibits a helical spin structure at low pressures \cite{Matsuda2016} and a quantum critical behavior around the critical pressure to the superconducting phase \cite{Cheng2015}.
The superconductivity in the vicinity of the helical magnetism is an interesting phenomenon to be explored \cite{Norman2015}. Therefore, it is important to compare the pressure dependence of the magnetic properties of CrAs and MnP in detail.
In particular, since MnP is ferromagnetic at ambient pressure, it is important to determine if the ferromagnetic interactions are still dominant and/or ferromagnetic fluctuations persist around the superconducting phase.
The pressure dependence of the magnetic structure in MnP was previously investigated up to 3.8 GPa using neutron diffraction technique \cite{Matsuda2016}, as shown in Sec. II A2.
Determining the magnetic structure near the superconducting phase requires challenging high pressure neutron diffraction measurements since the ordered magnetic moment is 0.84 $\mu\rm_B$ at 3.8 GPa and becomes even smaller at higher pressures and is expected to become very small around the critical pressure for the superconducting phase at 8 GPa. Although the diamond anvil cells are widely used in the pressure range, it is difficult to detect weak magnetic signals due to the small sample volume with an order of $\sim$0.1 mm$^3$ that can be loaded into the cell. In this work, we used a palm cubic anvil cell \cite{Dissanayake2019} to accommodate larger samples with an order of $\sim$1 mm$^3$ to detect weak magnetic Bragg peaks and successfully determined the magnetic structure up to \correction{7.5} GPa.

\begin{figure*}[htp]
	\centering
	\includegraphics[width=0.95\textwidth]{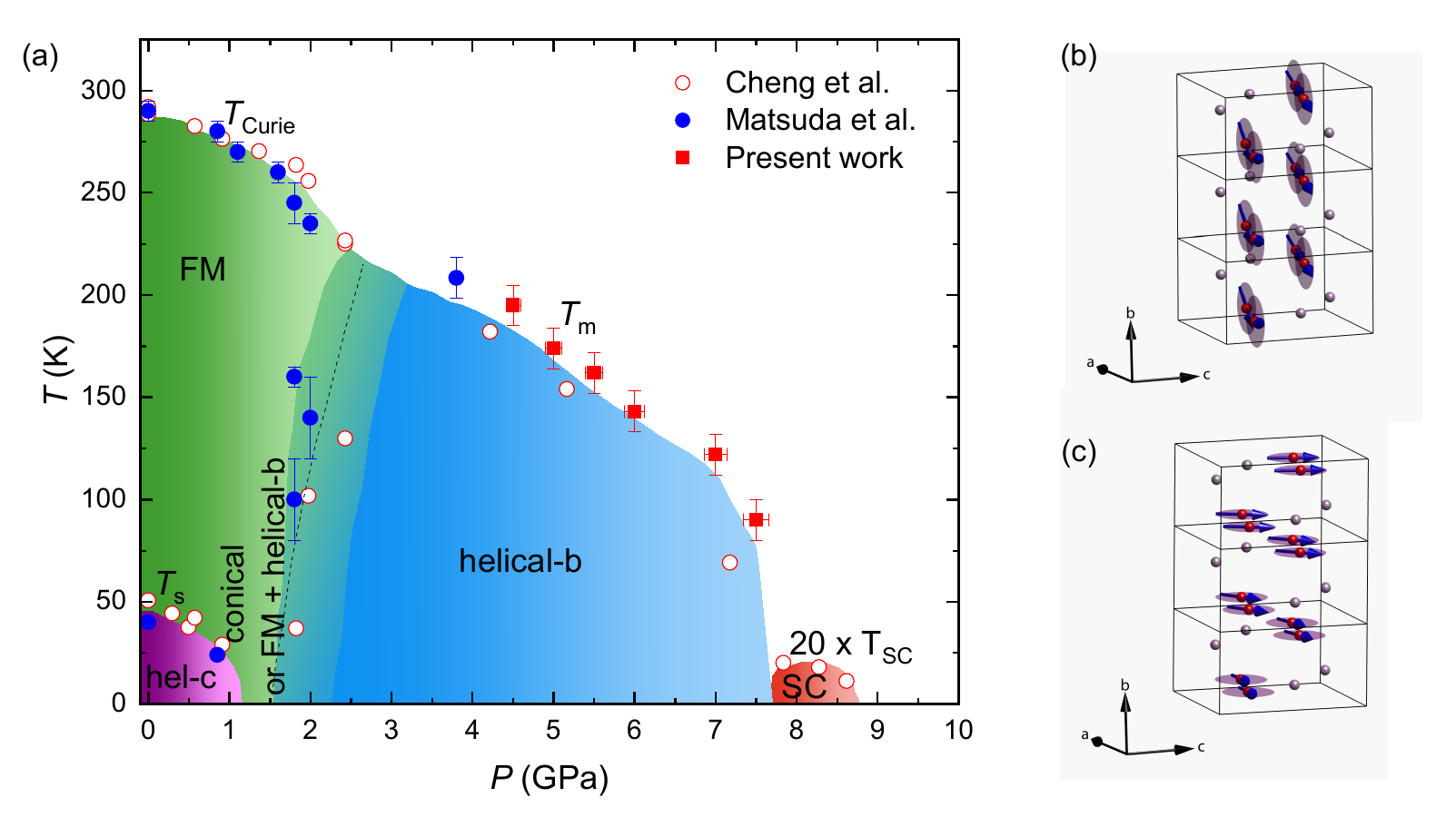}
	\caption{(a) The temperature-pressure phase diagram of MnP showing multiple magnetic phases and a superconducting phase. The open and filled circles are data from Cheng $et$ $al.$ (Ref. \cite{Cheng2015}) and Matsuda $et$ $al.$ (Ref. \cite{Matsuda2016}), respectively. The filled squares have been added in the present study. \correction{Note that the pressure is determined at room temperature and it gradually decreases by $\sim$0.5\%\ at 3 K.}
	(b) The magnetic structure in the ground state at low pressures $<$ 1.2 GPa (helical $c$). (c) The magnetic structure at high pressures $>$ 3 GPa (helical $b$). \correction{The red spheres with blue arrows are Mn atoms, whereas the purple spheres are P ions. } The shaded circles represent easy planes.
		\label{fig:phasediagram}
	}
\end{figure*}

Unfortunately, it is very difficult to determine the exchange interactions under high pressures using inelastic neutron scattering even with sizable single crystal. Therefore, theoretical analysis is used here to evaluate the exchange interactions.
Since multiple magnetic exchange paths are relevant to realize the ground states, theoretical analysis using first principles calculations and classical Monte Carlo simulations have been carried out. It was found that the two helical, ferromagnetic, and paramagnetic phases in Fig. \ref{fig:phasediagram}(a) are energetically close, and a transition between the two helical spin structures with applied pressure was reproduced qualitatively.
The results suggest that the exchange interactions change drastically with applied pressure, and while the near-neighbor interactions are predominant at low pressures, further-neighbor interactions become more influential at high pressures, where the superconductivity phase emerges. The theoretical study successfully reproduced the magnetic phase diagram consisting of multiple magnetic phases as a function of pressure, where a large number of exchange interactions are involved in causing the pressure-induced magnetic transitions.

The format of this paper is as follows. The magnetic structures and exchange interactions previously reported on materials with the MnP-type structures are described in Sec. II. Details of the high-pressure apparatus and the neutron and x-ray diffraction instruments are described in Sec. III. The pressure dependences of the magnetic structure, the magnetic transition temperature, and the magnetic moment are shown in Sec. IV. The theoretical analysis using first principles and classical Monte Carlo calculations is shown in Sec. V. The pressure dependence of the magnetic structure and the exchange interactions are discussed in Sec. VI. The results are summarized in Sec. VII.

\section{Previous results}
\subsection{magnetic structure}
\subsubsection{Ambient pressure}
The magnetic structures in materials with the B31 structure (also known as the MnP-type structure) have been extensively investigated. A double helical structure, in which a pair of helices is coupled with a finite spin angle between adjacent helical planes, was reported in MnP \cite{Felcher1966,Forsyth1966}, CrAs \cite{Boller1971,Selte1971}, FeP \cite{Felcher1971,Sukhanov2022}, and FeAs \cite{Selte1971,Rodriguez2011}. The orthorhombic structure in MnP can be described by either $Pnma$ ($b<a<c$) or $Pbnm$ ($c<b<a$) symmetries and both notations have been used in the literature. In this paper, we use the $Pnma$ notation and convert the notation from $Pbnm$ to $Pnma$ when referring to the literature using $Pbnm$.

MnP features two magnetically ordered states at ambient pressure: ferromagnetic order with the easy axis along the $b$ axis below 291 K , as well as double helical order with the magnetic propagation vector (0, 0, 0.112) and the easy $ab$ plane below 50 K (Fig. \ref{fig:phasediagram}(b))~\cite {Felcher1966,Forsyth1966}. The helical structure is not isotropic in the $ab$ plane but slightly elongated by 5\%\ along the $b$ axis
\cite{Moon1982}. We call this spin structure helical $c$ to distinguish from another helical structure at higher pressures described in Sec. II A2. A weak ferromagnetic component along the $a$ axis in the ferromagnetic phase and a slight tilt of the helical plane along the $c$ axis in the helical $c$ phase were reported in Ref. \cite{Yamazaki2014}. The successive magnetic transitions indicate that ferromagnetic interactions are dominant but frustrated antiferromagnetic interactions give rise to the helical magnetic structure at a low temperature.

CrAs shows a helical $c$ magnetic structure with the magnetic propagation vector (0, 0, 0.38) below 270 K. A characteristic feature is that the magnetic transition is of first order and accompanied by a large magnetoelastic coupling without changing the structural symmetry. The $b$ axis elongates by $\sim$5\%\ with a small contraction along the other axes.

The helical $c$ structure in FeP develops below 120 K with the magnetic propagation vectors (0, 0, 0.2). In FeAs a neutron powder diffraction study showed that a helical $c$ structure with the magnetic propagation vector (0, 0, 0.375) develops below 77 K \cite{Selte1971} and later a single crystal study with the polarized neutron technique revealed that the magnetic structure is better described with a noncollinear spin-density-wave (or elliptical helical structure) with the moment elongated along the $b$ axis \cite{Rodriguez2011}, as in MnP.

\subsubsection{Pressure dependence}
Matsuda $et$ $al.$ studied the pressure dependence of the magnetic structure in MnP \cite{Matsuda2016} using single-crystal neutron diffraction. The ferromagnetic transition temperature \correction{$T\rm_{Curie}$} decreases gradually with applied pressure. The helical transition temperature $T\rm_s$ also decreases with applied pressure and the helical $c$ phase finally disappears around 1.7 GPa. At 1.8 GPa, a ferromagnetic transition occurs at 250 K. Then the ferromagnetic component is reduced below $T\rm_m\sim$100 K accompanied by an emergence of another helical structure (helical $b$) with the magnetic propagation vector (0, $\delta$, 0) and the easy $ac$ plane (Fig. \ref{fig:phasediagram}(c)). As in the helical $c$ structure, the helical $b$ structure is elliptical with the moment along the $a$ axis elongated. Therefore, the ground state structure at 1.8 GPa is a conical structure or a mixed phase of the ferromagnetic and helical $b$ structures. This magnetic structure is consistent with that obtained from $\mu$SR measurements \cite{Khasanov2016}. A neutron powder diffraction study also showed \correction{a broad} incommensurate magnetic peak above 2 GPa that is consistent with the helical $b$ structure, although it showed another incommensurate magnetic structure at 1.5 GPa with the propagation vector (0.25, 0.25, 0.125) \cite{Yano2018}, which might be stable in a very narrow pressure range. 
With increasing pressure, \correction{$T\rm_{Curie}$} decreases and $T\rm_m$ increases. Above $\sim$2.5 GPa, the ferromagnetic phase or component disappears and the helical $b$ structure was found to be the magnetic ground state at 3.8 GPa \cite{Matsuda2016}. The temperature-pressure phase diagram is shown in Fig. \ref{fig:phasediagram}(a).

Pressure dependence of the magnetic structure was also studied using x-ray magnetic diffraction up to 8.99 GPa \cite{Wang2016}. Incommensurate peaks were observed at (0, 0, 1$\pm\delta$) with $\delta\sim$0.25 at 3.17, 5.28, and 6.43 GPa, suggesting that another type of helical $c$ structure appears at higher pressures above 3 GPa. However, neutron scattering experiments have not confirmed this helical phase.
One difference between Ref. \cite{Wang2016} and the other experiments was the different thermodynamical path. The pressure was applied at a low temperature in Ref.~\cite{Wang2016}, whereas the pressure was applied at room temperature in other measurements~\cite{Matsuda2016,Khasanov2016,Yano2018}.
Since this result is very different from that obtained by the neutron and muon studies \cite{Matsuda2016,Khasanov2016,Yano2018}, the pressure dependence of the magnetic ground state must be evaluated with further studies.

As reported in Ref. \cite{Wu2014}, the magnetic transition temperature in CrAs decreases gradually with increasing pressure and the helical $c$ magnetic phase disappears around 0.8 GPa. The magnetic structure as a function of pressure was investigated using neutron diffraction \cite{Shen2016}. The magnetic propagation vector (0, 0, $\delta$) changes with changing temperature and pressure. $\delta$ decreases with decreasing temperature and increasing pressure. An interesting feature is that reorientation of the helical plane occurs from the $c$ plane to the $b$ plane around 0.6 GPa, where bulk superconductivity emerges, without changing the direction of the magnetic propagation vector.

In CrAs, chemical pressure can be applied by substituting P for As \cite{Suzuki1993,Kanaya2004}. Physical properties were studied in the chemical pressured system CrAs$_{1-x}$P$_x$ \cite{Matsuda2018}. The quantum critical behavior was observed as for the physical pressure. The pressure dependence of the magnetic structure was also similar to that for the physical pressure. The only difference is that the phosphorus doped system shows no superconductivity around the quantum critical region probably due to the disorder caused by the doping.
Since such chemical pressure is not possible in MnP, pressure effects should be studied only with physical pressures.

\subsection{Exchange interactions}
\begin{figure*}[tb]
	\centering
	\includegraphics[width=0.8\textwidth]{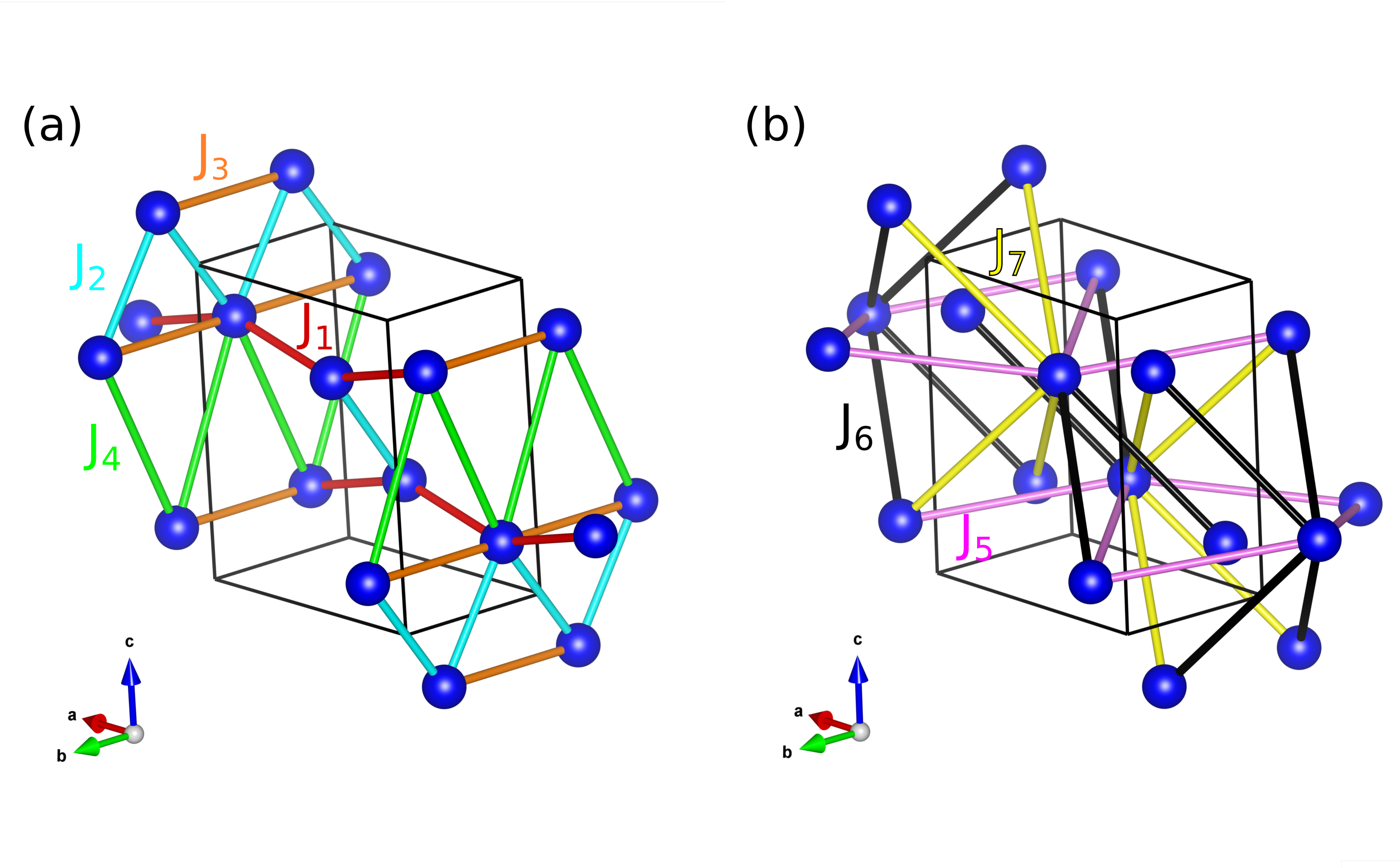}
	\caption{Exchange interactions \correction{$J_1$, $J_2$, $J_3$, and $J_4$ (a) and $J_5$, $J_6$, and $J_7$ (b)} in the MnP-type structure. Only magnetic ions are depicted.
	\label{interactions}
	}
\end{figure*}

Previous inelastic neutron scattering measurements and theoretical calculations have determined that a substantial number of exchange paths are needed to describe the magnetism in MnP-type compounds~\cite{Kallel1974,Takeuchi1968,Dobrzynski1989,Sukhanov2022,Tajima1980,Todate1987,Yano2012,Yano2013,Itoh2014,Xu2017}, as depicted in Fig. \ref{interactions}.
Table \ref{distance} lists distance for each exchange interaction in MnP, CrAs, and FeP. There is a gap of the distance between $J_3$ and $J_4$. However, $J_4$ should be included in CrAs, and even further-neighbor interactions are important in MnP, as described below.

Kallel $et$ $al.$ \cite{Kallel1974} proposed an exchange interaction model with three interactions to explain the magnetic structures in CrAs, MnP, and FeP. According to the model ($J_1$-$J_2$-$J_4$ model), the competing interactions between $J_1$, $J_2$, and $J_4$, where $J_n$ represents the magnetic coupling between the $n$-th neighbor spins, give rise to the helical structure.
The helical structure in CrAs can be reproduced with $J_4$/$J_1$=$-$0.52 and $J_2$/$J_1$=7.1, where $J_1$ and $J_2$ are antiferromagnetic and $J_4$ is ferromagnetic, suggesting that antiferromagnetic $J_2$ is the dominant interaction and competes with $J_1$ and $J_4$.
In MnP, other theoretical studies \cite{Takeuchi1968,Dobrzynski1989} reported that interactions up to the seventh neighbors are needed to reproduce the helical magnetic structure.

\begin{table}
\caption{Distance between magnetic ions for each exchange interaction shown in Fig. \ref{interactions}, calculated using the structural parameters at room temperature for MnP (Ref.~\cite{Forsyth1966}), CrAs (Ref.~\cite{Boller1971}), and FeP (Ref.~\cite{Felcher1971}). The unit is \AA. It is noted that the distance for $J_5$ is slightly larger than that for $J_6$ in CrAs.}
\label{distance}
\begin{ruledtabular}
\begin{tabular}{cccccccc}
 & $J_1$ & $J_2$ & $J_3$ & $J_4$ & $J_5$ & $J_6$ & $J_7$\\
\hline
MnP & 2.704 & 2.816 & 3.172 & 3.927 & 4.168 & 4.233 & 4.296\\
CrAs & 2.889 & 3.011 & 3.445 & 4.111 & 4.496 & 4.453 & 4.607\\
FeP & 2.657 & 2.786 & 3.093 & 3.805 & 4.078 & 4.152 & 4.217\\
\end{tabular}
\end{ruledtabular}
\end{table}

In FeP, Sukhanov $et$ $al.$ observed magnetic excitations using inelastic neutron scattering technique \cite{Sukhanov2022}. They first evaluated the $J_1$-$J_2$-$J_4$ model, which they call the frustrated trapezoid model, and found that the observed magnetic excitations cannot be reproduced by the model. They observed a steep dispersion extending up far above 50 meV along $k$ direction, which indicates strong ferromagnetic interactions along the $b$ axis. A model of zigzag interchain interactions along the $c$ axis with additional small interactions along the $a$ axis is then proposed to reproduce the magnetic dispersions.

Inelastic neutron scattering studies have been reported in MnP \cite{Tajima1980,Todate1987,Yano2012,Yano2013,Itoh2014}. An isotropic Heisenberg model with up to 6th-neighbor interactions was fitted to reproduce the experimental magnetic dispersions, and it was reported that $J_2$ ($-$0.65 meV) and $J_4$ ($-$0.06 meV) are antiferromagnetic, whereas $J_1$ (0.37 meV), $J_5$ (0.26 meV), and $J_6$ (0.64 meV) are ferromagnetic \cite{Yano2013,Itoh2014}.
The ferromagnetic interactions were reported to be dominant in total and competing antiferromagnetic interactions $J_2$ and $J_4$ were suggested to give rise to the helical $c$ structure at low temperatures. Although $J_3$ has not been determined so far, the magnetic dispersion is less steep along the $b$ axis than along the other directions \cite{Todate1987}, suggesting that $J_3$ is not large as in FeP described above.
Those previous inelastic neutron scattering studies indicate that the experiments are challenging even at ambient pressure probably due to the small Mn moments and large excitation energies ($\sim$70 meV). Therefore, observing the full magnetic dispersion relations under high pressures is not currently feasible.

More recently, first-principles calculations have been carried out to examine the exchange interactions and the spin structure in MnP at ambient pressure with sharply contrasting results compared to those mentioned above. Tran $et$ $al$. reported $J_1$, $J_2$, and $J_3$ to be ferromagnetic with absolute values larger than 10 meV, $J_4$ and $J_5$ to be antiferromagnetic with absolute values $\sim 5$ meV, and $J_6$ and $J_7$ to be slightly ferromagnetic at $\sim 2$ meV \cite{Tran2021}. Interactions beyond 7th nearest neighbor were reported to be smaller. Reproduction of the helical magnetic structure was not attempted, and it remains to be seen if these $J$ values can reproduce the experimentally determined magnetic structures. 
On the other hand, Xu $et$ $al$. fitted first-principles results assuming the $J_1$-$J_2$-$J_4$ model and successfully reproduced the helical to ferromagnetic transition with increasing pressure \cite{Xu2017}. However, the interaction magnitudes were predicted to be as large as 100 meV, which is too large considering the experimental paramagnetic transition temperatures. It is also noted that they successfully reproduced the pressure-induced transitions from helical $c$ to ferromagnetic and to helical $b$ structures at 0 K using noncollinear density functional theory calculations but were unable to provide interpretations in terms of interaction parameters.

\section{Experimental Details}
MnP single crystals used in this study were grown by the Bridgman method from stoichiometric mixtures of Mn and P. A palm cubic anvil cell, which consists of a cluster of six ZrO$_2$ anvils converging onto the center gasket from three orthogonal directions, was used to apply hydrostatic pressures up to \correction{7.5} GPa \cite{Dissanayake2019}. Single crystal samples of 2$\times$2$\times$2 mm$^3$ and 0.9$\times$0.9$\times$0.9 mm$^3$ were used in 4 mm (for $P$=4.5 GPa) and 2.5 mm (for $P >$4.8 GPa) anvil setups, respectively. Fluorinert was used as the pressure-transmitting medium. High pressure single crystal diffraction experiments were performed on HB-1 and HB-3 thermal triple axis spectrometers at High Flux Isotope Reactor (HFIR) at Oak Ridge National Laboratory (ORNL).
Fixed incident neutron energies of 13.5 meV and 14.7 meV were used on HB-1 and HB-3, respectively. The horizontal collimator sequence was 48’-80’-sample-80’-240’ and the contamination from higher-order beams was eliminated using Pyrolytic Graphite (PG) filters. A high capacity closed cycle refrigerator (CCR) was used to cool the sample down to 3 K. Magnetic structure refinements were performed using Fullprof \cite{RODRIGUEZCARVAJAL1993} and estimated the magnetic moment at 3 K. 

 The high pressure single crystal x-ray diffraction experiments were performed at the 4-ID-D beamline at the Advanced Photon Source (APS) at the Argonne National Laboratory. The incident energy of the x-rays was 20 keV and a Sumitomo CCR was used to cool down the diamond anvil cell to 4 K.  The single crystals of MnP are cut in to crystals with dimensions of 60 $\mu$m x 60 $\mu$m x 40 $\mu$m. The sample quality and initial orientation were checked by Laue x-ray diffraction. For high pressure x-ray experiment,  Boehler-Almax type 800 $\mu$m culet diamond anvil cell was used and 4:1 mixture of Methanol:Ethanol was used as the hydrostatic pressure medium. A  gold foil was also loaded as the manometer. The pressure was determined by the equation of state of Au~\cite{Holzapfel2001}. A helium gas membrane was used to control the in-situ pressure.
 
\section{Experimental Results}

\begin{figure*}[t]
	\centering
	\includegraphics[width=0.95\textwidth]{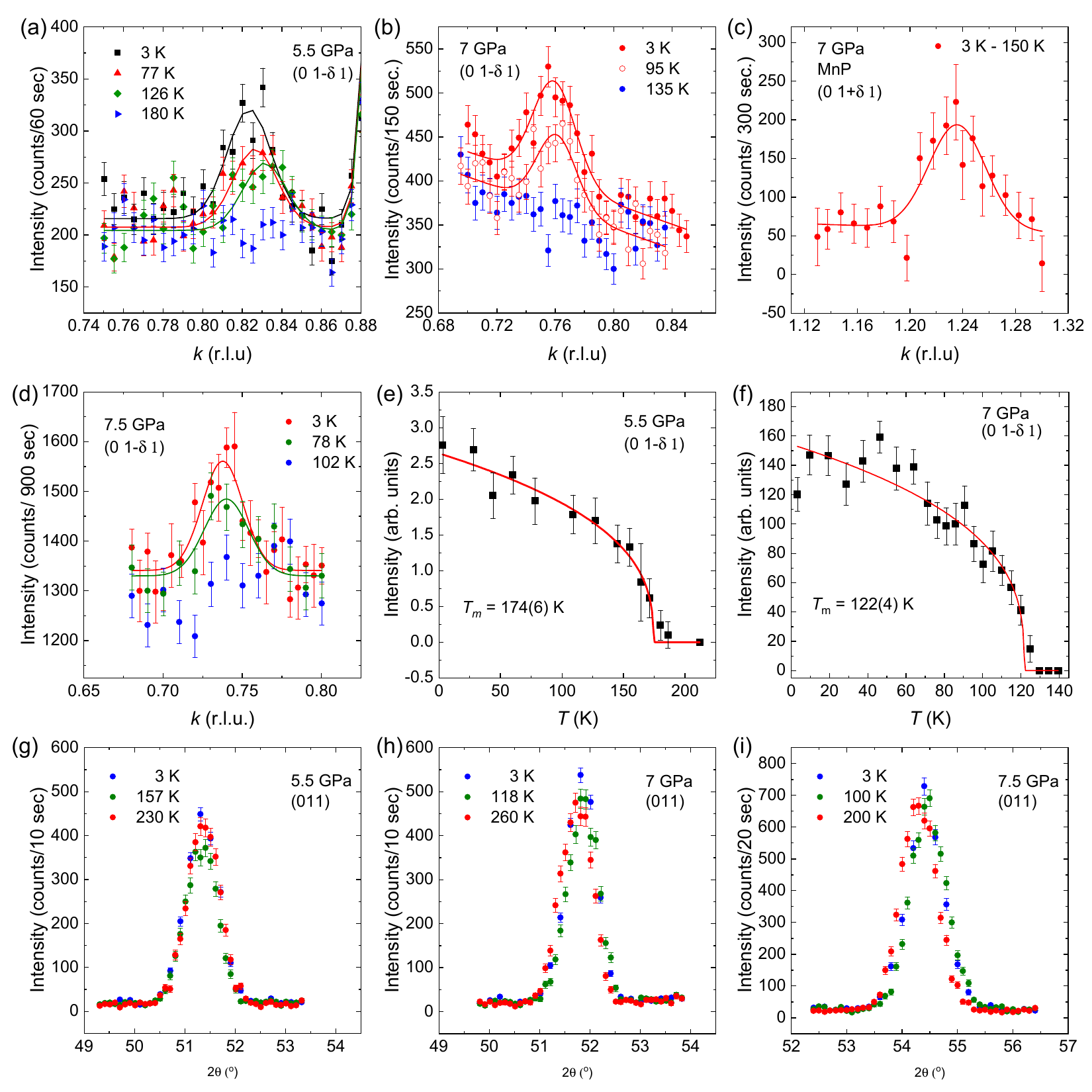}
	\caption {\correction{Incommensurate magnetic Bragg intensioty at (0 $1-\delta$ 1) measured at (a) 5.5 GPa, (b) 7 GPa, and (d) 7.5 GPa and at (0 1+$\delta$ 1) measured at (c) 7 GPa.}
    The intensities in \correction{(a), (b) and (d)} are raw data, whereas those in (c) are the data at 3 K subtracted the 150 K data as a background. Note that a constant background remains due to the temperature dependent signal from the pressure cell. The solid lines are fits to a Gaussian function.
	Temperature dependence of the (0 1-$\delta$ 1) magnetic Bragg peak intensity measured at \correction{(e)} 5.5 GPa and \correction{(f)} 7 GPa. The solid lines are the fitting results to a power-law function.
    \correction{$\theta$-2$\theta$ scans of the (011) nuclear Bragg peak at 3 K, just below $T\rm_{m}$, and a temperature higher than 200 K measured at (g) 5.5 GPa, (h) 7 GPa, and (i) 7.5 GPa.
    The measurements at 5.5 and 7 GPa were performed on HB-3 and the measurements at 7.5 GPa were performed on HB-1.}
		\label{fig:magneticpeaks}
	}
\end{figure*}

\begin{figure*}[t]
	\centering
	\includegraphics[width=0.95\textwidth]{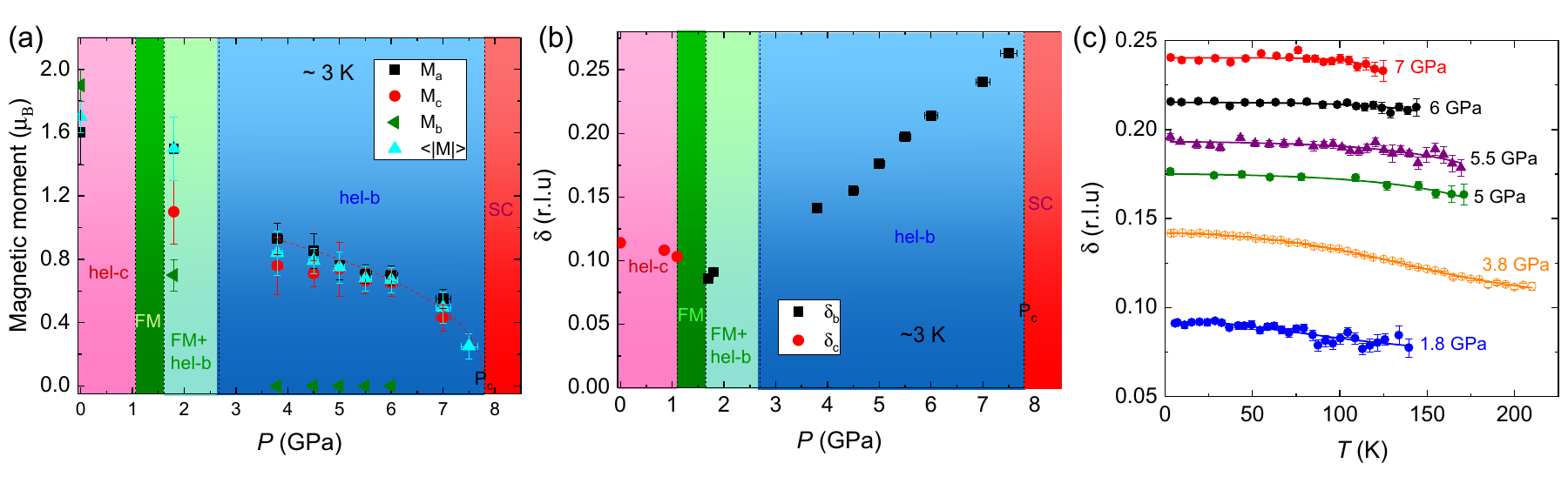}
	\caption{(a) Magnetic moment M$_a$, M$_b$, and M$_c$ and magnitude of the average moment $<|M|>$ around 3 K as a function of applied pressure. The broken line is a guide to the eye. (b) Incommensurability, $\delta$ as a function of pressure for each different magnetic ground state. The different background colors correspond to different magnetic ground states in the phase diagram. \correction{Note that the y-error bars are smaller than the symbols.} (c) The temperature dependence of $\delta$ as a function of pressure. The solid lines are guides to the eye.
	\label{fig:moment_delta_P_dep}
	}
\end{figure*}

\begin{figure*}[t]
	\centering
	\includegraphics[width=0.95\textwidth]{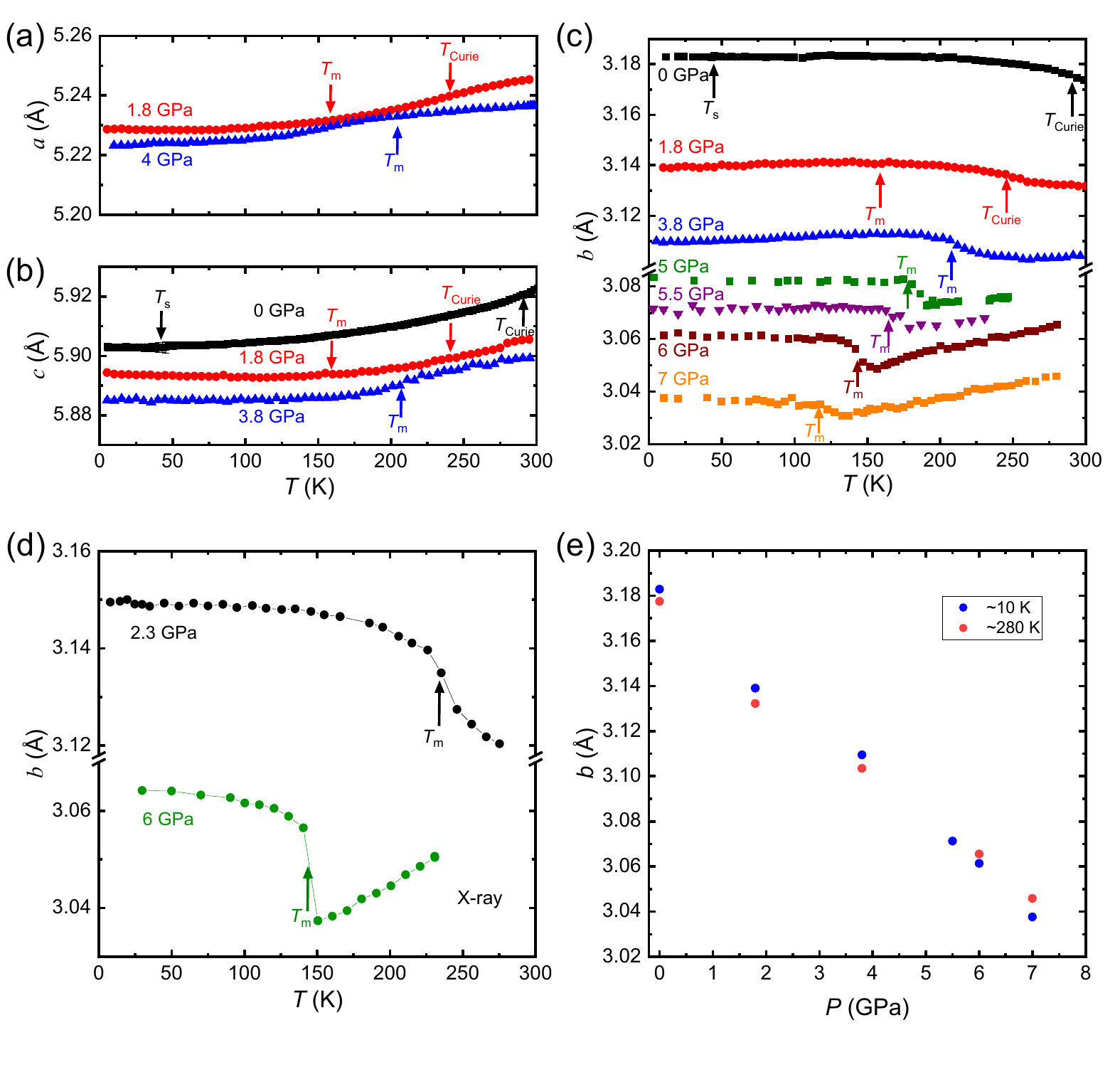}
	\caption{(a), (b), and (c) Temperature dependence of the lattice constants $a$, $c$, and $b$, respectively, measured using neutron diffraction technique as a function of pressure labeled in each panel.
	(d) Temperature dependence of the lattice constant $b$ at 2.3 and 6 GPa measured using synchrotron x-ray diffraction technique. The solid lines are guides to the eye.
	\correction{The ferromagnetic transition temperature $T\rm_{Curie}$, the helical $c$ transition temperature $T\rm_s$, and the helical $b$ transition temperature $T\rm_m$ are marked with arrows. The transition temperatures are derived from the neutron diffraction results shown in Fig. \ref{fig:phasediagram}(a). Interpolated values are used for the data in (d). (e) Pressure dependence of the lattice constant $b$ at $\sim$10 and $\sim$280 K.}
	\label{fig:neutron_lattice_const}
    }
\end{figure*}

In order to investigate the magnetic structure in the vicinity of superconducting phase, high pressure neutron diffraction measurements were successfully performed with the palm-cubic anvil cell up to \correction{7.5} GPa, which are challenging pressures for magnetic diffraction studies.
\correction{Details of this pressure cell, including the pressure dependence of a nuclear Bragg reflection in MnP, are described in Ref. \cite{Dissanayake2019}.}

Figures \ref{fig:magneticpeaks}(a)-\ref{fig:magneticpeaks}(d) exhibit typical magnetic Bragg peak profiles observed in the pressure range between 5.5 and \correction{7.5} GPa.
\correction{For reference, the temperature dependence of the (011) nuclear Bragg peak in the same pressure range is shown in Figs. \ref{fig:magneticpeaks}(g)-\ref{fig:magneticpeaks}(i). The peak position shifts gradually to higher angles with appplied pressure and the peak width is almost unchanged up to 7.5 GPa, indicating that the pressure is homogeneous and does not induce lattice distortion.}
We observed incommensurate magnetic peaks at (0, 1-$\delta$, 1) and (0, 1+$\delta$, 1) corresponding to the helical $b$ structure previously observed at 3.8 GPa \cite{Matsuda2016}. The peak width is close to resolution-limited in the whole pressure range so that the magnetic order is considered to be long-ranged. This indicates that the helical $b$ phase is stable as the magnetic ground state in a wide pressure range up to \correction{7.5} GPa.
The peak position moves away from (0, 1, 0) gradually with increasing pressure and also cooling down, indicating that the incommensurability $\delta$ increases at higher pressures and lower temperatures. The pressure and temperature dependences of $\delta$ are plotted in Fig. \ref{fig:moment_delta_P_dep}(c), which shows the aforementioned pressure and temperature dependences.
We also searched for Bragg peaks at (0,0,1$\pm\delta$) and the equivalent positions, since Ref. \cite{Wang2016} reported the magnetic peaks above 3 GPa. However, we did not observe any peaks at those positions.
\correction{We suspect that the incommensurate peaks observed at high pressures in the x-ray study~\cite{Wang2016} might be very weak signal, which originates from a minor phase and is not observable with neutron diffraction, or sample dependent signal or due to surface effect.}
We conclude that the helical $b$ is the accurate magnetic phase adjacent to the superconducting phase.
The order parameter of the integrated magnetic intensities of (0 1-$\delta$ 1) magnetic Bragg peak for 5.5 and 7 GPa are shown in Figs. \ref{fig:magneticpeaks}(e) and \ref{fig:magneticpeaks}(f), respectively. The transition temperatures agree with the previously reported phase diagram. The magnetic structure refinements were performed for all the pressure points from 0 to \correction{7.5} GPa using Fullprof program for corresponding helical $c$ ($P <$ 1.3 GPa) and helical $b$ ($P >$ 1.8 GPa) phases.
The updated phase diagram with the present results is shown in Fig. \ref{fig:phasediagram}(a).
\correction{We found a long range magnetic ordered phase in the vicinity of the superconducting phase. On the other hand, the high pressure measurements with a powder sample by Yano $et$ $al.$ did not show magnetic signal around 7 GPa~\cite{Yano2018}. This might be due to extremely weak magnetic signal from a powder sample with magnetic moments of 0.3$\mu\rm_B$ around 7 GPa.}

The pressure dependence of the ordered magnetic moments along each direction $M_a$, $M_b$ and $M_c$ and the magnitude of the average magnetic moment $<|M|>$ at $T$=3 K are shown in Fig. \ref{fig:moment_delta_P_dep}(a). We observe that the magnitude of the ordered magnetic moment monotonically decreases with applied pressure, indicating that the itinerancy of the Mn moments is enhanced with applied pressure. The trend of the pressure dependence of the ordered magnetic moment up to \correction{7.5} GPa suggests that static magnetic ordering completely disappears at the critical pressure $P_c$ where the superconductivity dome appears.
The pressure dependence of $\delta$ at $T$=3 K for helical $c$ and helical $b$ phases are shown in Fig. \ref{fig:moment_delta_P_dep}(b). The $\delta$ of helical $c$ phase decreases with pressure and the $\delta$ of helical $b$ phase linearly increases with pressure.

The temperature dependence of the lattice constants measured using neutron diffraction experiments are shown in Figs. \ref{fig:neutron_lattice_const}(a)-\ref{fig:neutron_lattice_const}(c). The temperature dependence of the lattice constants $a$ and $c$ show a normal behavior, which contract with decreasing temperature. Anomalies in the lattice constants $a$ and $c$ are negligible at the magnetic transition temperatures. In contrast, the lattice constant $b$ exhibits an anomalous behavior. It expands below \correction{$T\rm_{Curie}$} and $T\rm_m$ and stays almost constant down to low temperatures.
This anomaly becomes sharper at higher pressures up to 6 GPa and then is reduced at 7 GPa. Note that the $b$ axis expands both at the transition to the ferromagnetic and helical $b$ phases and no anomaly is observed at $T\rm_s$ at ambient pressure and at $T\rm_m$ at 1.8 GPa.

\correction{The pressure dependence of the lattice constant $b$ is shown in Fig. \ref{fig:neutron_lattice_const}(e). As reported in Refs. \cite{Yano2018,Wang2016}, it is contracted significantly with applied pressure. At lower pressures below 4 GPa, the $b$ axis becomes larger at $\sim$20 K than at $\sim$280 K due to the lattice expansion along the $b$ axis at $T\rm_{m}$. On the other hand, at higher pressures above 5.5 GPa, it becomes smaller at $\sim$20 K than at $\sim$280 K in spite of the lattice expansion at $T\rm_{m}$.
Temperature dependence of the (011) nuclear Bragg peak measured at 5.5, 7, and 7.5 GPa is shown in Figs. \ref{fig:magneticpeaks}(g)-\ref{fig:magneticpeaks}(i). The peak position is almost temperature independent at 5.5 GPa, whereas the peak shifts to higher scattering angles with decreasing temperature at 7 and 7.5 GPa. This is consistent with the pressure dependence of the lattice constant $b$ mentioned above.}

To investigate the crystal structure in more detail as a function of pressure and temperature, synchrotron x-ray diffraction experiments were performed using a single crystal.
The crystal structure remains same as the ambient pressure structure with $Pnma$ symmetry with no evidence for a structural transition with pressure. An anomaly of the lattice constant $b$ at $T\rm_m$ 
similar to that observed using neutron diffraction
was observed in the x-ray diffraction experiments, as shown in Fig. \ref{fig:neutron_lattice_const}(d). A very sharp increase of the lattice constant $b$ of about 1.15 \% was observed at 6 GPa at $T\rm_m$.

\section{Theoretical Analysis}
To capture the trend of structural stability and magnetization interaction with respect to pressure from a theoretical point of view, we performed first-principles calculations using the software package VASP \cite{Kresse1996,Kresse1996a} and extracted the interactions using wannier90 \cite{Pizzi2020} and TB2J \cite{He2021}. 
First, we used VASP to perform first-principles calculations using the GGA-PBE exchange correlation functional \cite{Perdew1996,Perdew1996a} and the projector augmented wave method \cite{Bloechl1994} to represent ion-electron interactions. A plane wave energy cutoff of 400 eV was employed for the wave function expansion, and a 6 $\times$ 10 $\times$ 6 $\mathbf{k}$-point mesh was used for Brillouin zone sampling. The unit cell shown in Fig.~\ref{interactions} with 4 MnP units were used for the calculations. \correction{We fixed the lattice constants to the experimentally reported ones \cite{Wang2016} at low temperature in the pressure range of 0 to 10.29 GPa and relaxed the internal parameters}. A collinear ferromagnetic structure was assumed in the first-principles calculations. A tight-binding model based on maximally localized Wannier functions corresponding to Mn $d$ and P $s$ orbitals were extracted from the calculated electronic structure using wannier90 code, then fed into the TB2J code to calculate the exchange interactions based on the Liechtenstein formalism \cite{Liechtenstein1987,Korotin2015,He2021}. The exchange interactions thus obtained were used for Monte Carlo simulations of the Heisenberg model 
\begin{equation}
    H = -\sum_{ij} J_{ij} \mathbf{S}_{i} \cdot \mathbf{S}_{j}
\end{equation}
to examine the possibility of helical magnetic structures using VAMPIRE code \cite{Evans_2014}.

\begin{figure}[t]
	\centering
	\includegraphics[width=0.4\textwidth]{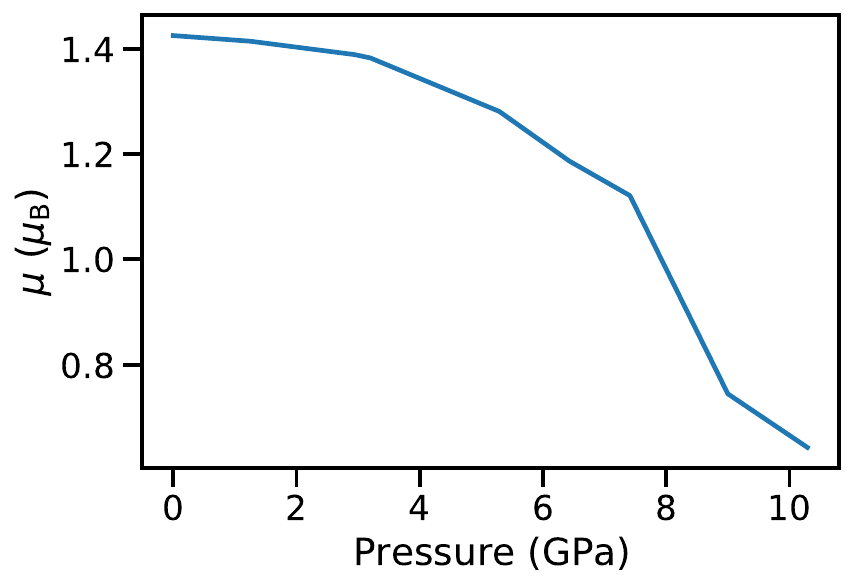}
	\caption{The calculated pressure dependence of the magnetic moment per MnP unit in the collinear ferromagnetic structure.
	\label{fig:DFT_magmom}
	}
\end{figure}

\begin{figure}[t]
	\centering
	\includegraphics[width=0.45\textwidth]{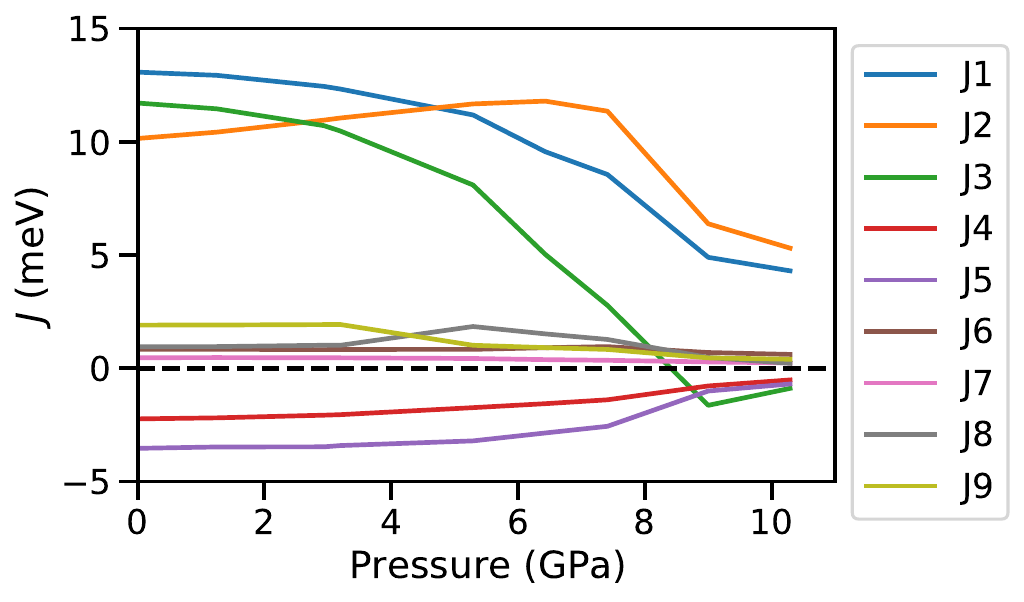}
	\caption{The calculated pressure dependence of the exchange interactions $J$ up to the ninth nearest neighbor. Positive and negative values correspond to ferromagnetic and antiferromagnetic interactions, respectively.
	\label{fig:theor_J}
	}
\end{figure}

\begin{figure}[t]
	\centering
	\includegraphics[width=0.5\textwidth]{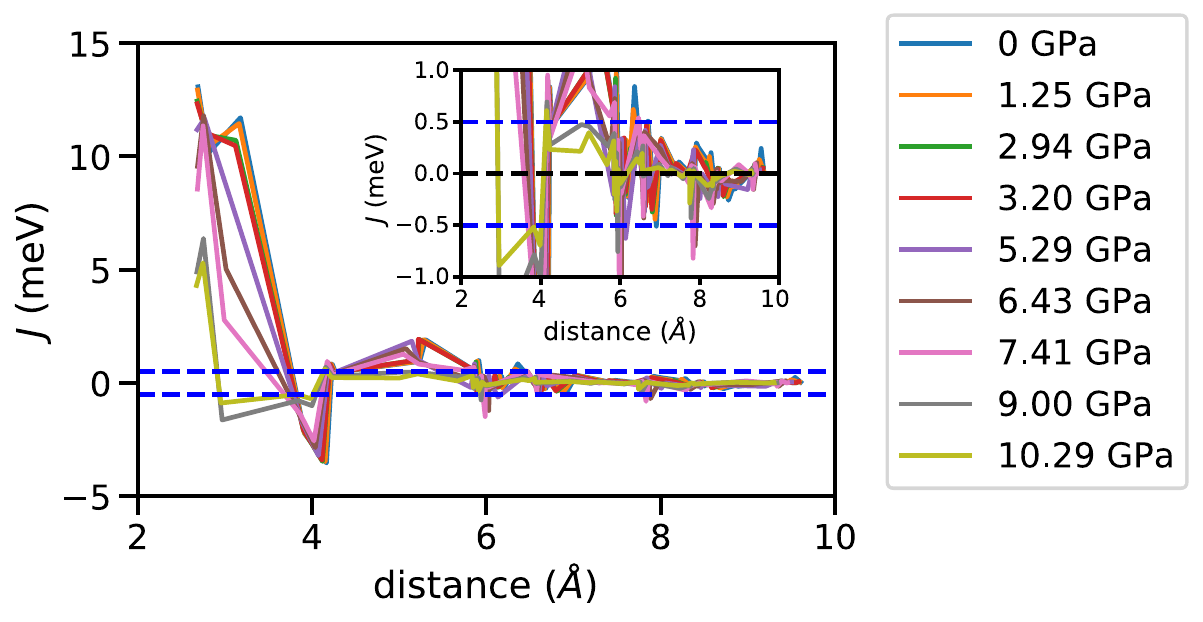}
	\caption{The calculated exchange interactions $J$ up to the 51st nearest neighbor plotted vs.~bond distance. Positive and negative values correspond to ferromagnetic and antiferromagnetic interactions, respectively. Dashed horizontal lines are drawn at $\pm 0.5$ meV as guides for the eye.  The inset shows an expanded figure in the range of $-1 \text{ meV} < J < 1 \text{ meV}$.
	\label{fig:theor_JR}
	}
\end{figure}

The magnetic moments calculated in the collinear ferromagnetic configuration are presented in Fig.~\ref{fig:DFT_magmom}. Although magnetic moments are suppressed with pressure, a finite moment persists within the calculated pressure range.

\begin{figure*}[htbp]
	\centering
	\includegraphics[width=0.9\textwidth]{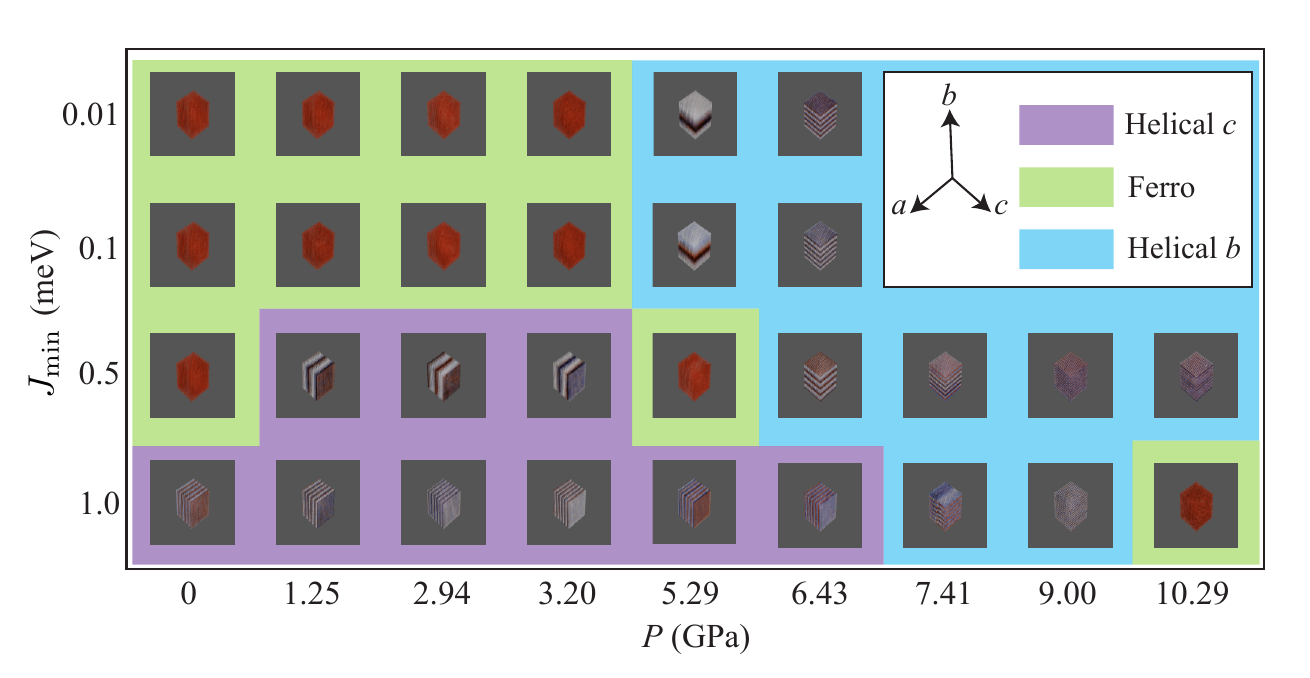}
	\caption{The snapshots of magnetic structures obtained at $1.8\times 10^5$ MC step under the pressure $P$ = 0, 1.25, 2.94, 3.20, 5.29, 6.43, 7.41, 9.00, and 10.29~GPa for $J_\mathrm{min}$ = 0.01, 0.1, 0.5, and 1 meV, respectively. 
    Red and blue represent spins with $S_z$ components of $1$ and $-1$ along the $c$-axis, respectively~\cite{vdc}. 
	\label{fig:theor_alpha_dep}
	}
\end{figure*}
\begin{figure*}[htb]
	\centering
	\includegraphics[width=0.9\textwidth]{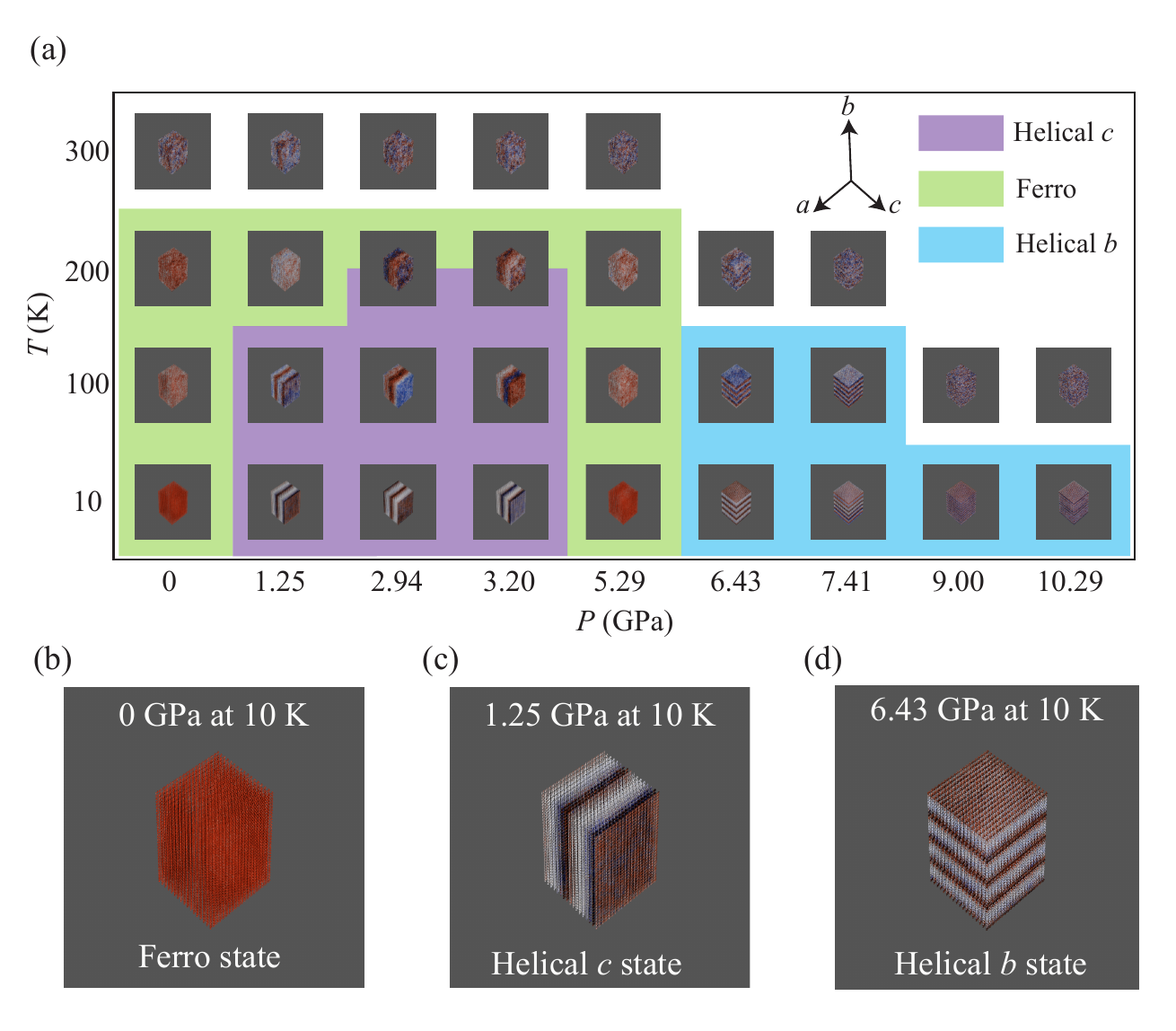}
	\caption{(a) The snapshots of magnetic structures obtained at $1.8\times 10^5$ MC step for $T$ = 10, 100, 200, and 300~K under the pressure $P$ = 0, 1.25, 2.94, 3.20, 5.29, 6.43, 7.41, 9.00, and 10.29~GPa with $J_\mathrm{min} = 0.5$ meV. \correction{The enlarged figures of the snapshots of magnetic structures for (b) $P=0$ GPa, (c) $P=1.25$ GPa, and (d) $P=6.43$ GPa at $T=10$ K.} Red and blue represent spins with $S_z$ components of $1$ and $-1$ along the $c$-axis, respectively~\cite{vdc}. 
	\label{fig:theor_phase}
	}
\end{figure*}
The extracted exchange interactions between Mn atoms are shown as functions of pressure in Fig.~\ref{fig:theor_J} up to the ninth nearest neighbor and as functions of distance for varying pressures in Fig.~\ref{fig:theor_JR}. Interactions involving P are ignored here as their magnitudes are negligible in comparison to Mn-Mn interactions. At 0 GPa, the $J_1$, $J_2$, and $J_3$ interactions are all found to be ferromagnetic with values higher than 10 meV at 0 GPa while the $J_4$ and $J_5$ interactions are antiferromagnetic at $\sim -2$ meV and $-3.5$ meV, respectively. The $J_6$, $J_7$, $J_8$, and $J_9$ are slightly ferromagnetic. \correction{These values are in good agreement with a previous theoretical work employing the all-electron KKR approach \cite{Tran2021} but disagree with several other reports \cite{Yano2013,Itoh2014,Xu2017}; we return to this point in Sec.~\ref{sec:discussion}.}

With increasing pressure, the $J_1$ and $J_3$ interactions decrease in magnitude while $J_2$ increases in magnitude and becomes the strongest interaction above $\sim 5$ GPa. Above $\sim 7$ GPa, the $J_2$ also starts to decrease. 
$J_1$ and $J_2$ both stay ferromagnetic up to 10 GPa although the interactions are suppressed down to $\sim 5$ meV and $\sim 6$ meV, respectively. In comparison, the antiferromagnetic $J_4$ and $J_5$ interactions as well as the slightly ferromagnetic $J_6$ and $J_7$ interactions depend little on pressure. 

The most notable change with pressure is seen for the $J_3$ interaction: it changes from a ferromagnetic interaction ($J_3 = 11.7$ meV) to an antiferromagnetic interaction ($J_3 = -1.6$  meV) when the pressure is increased from 0 GPa to 9 GPa. 
The $J_3$ interaction is in the $b$-axis direction (Fig.~\ref{interactions}), so its quick suppression with pressure compared to other interactions is in accordance with the anisotropic pressure dependence of the lattice constants \cite{Wang2016,Yano2018}: the $b$ axis is contracted by $\sim$5.5\%\ between 0 and 7 GPa, whereas the $a$ and $c$ axes show a minor change within 0.4\% in the same pressure range. 

It is also noted that interactions larger than 0.5 meV exist beyond the 10th nearest neighbor, corresponding to $\sim$5.9 \AA, as shown in Fig.~\ref{fig:theor_JR}. We show below that \correction{the anisotropic dependence of the $J$ values on the bonding direction and} these long-range interactions actually contribute to the phase transition behavior, and that shorter-ranged models proposed in previous works may be too simple to describe the observed magnetic phases vs.~pressure.

Using the obtained magnetization interactions for the Heisenberg model, we performed classical Monte Carlo (MC) calculations of temperature- and pressure-dependent spin configurations using the numerical code \verb|VAMPIRE|. The calculation cell size was set to the largest supercell dimension that fits within a (10 nm)$^3$ cubic box. We performed calculations where $J$ parameters below thresholds $J_\mathrm{min}$ of $0.01$, $0.1$, $0.5$, and $1$ meV were set to zero.  This was done to examine the convergence of the obtained configurations with the number of exchange interactions, but as we discuss below, these calculations were also very useful for elucidating the interactions that stabilize the various phases observed in experiment. We performed $2 \times 10^5$ MC steps for each case and confirmed that all calculations are well converged above $1.5\times 10^5$ MC steps.

Figure \ref{fig:theor_alpha_dep} shows the pressure dependence of spin structures with $J_\mathrm{min}=0.01, 0.1, 0.5, $ and $1$ meV at $10$ K, respectively. Here, the spin structure snapshots obtained at $1.8 \times 10^5$ MC steps are displayed.
Ideally, the result would converge to the experimentally observed phases with decreasing $J_\mathrm{min}$, but this is not the case, as the ferromagnetic phase is wrongly predicted to be stable at 0 K and from $0$ to $3.20$ GPa for $J_\mathrm{min}=0.1$ meV and lower. This may be because of numerical errors as well as the various approximations in the theory and calculation procedure including neglect of Dzyaloshinskii–Moriya interactions, single ion anisotropy, temperature-dependence of the exchange interactions, and interaction with the conduction electrons. On the other hand, all of the experimental phases near $0$ K are reproduced rather well when $J_\mathrm{min}$ is $0.5$ meV with ferromagnetic, helical $c$, ferromagnetic, and helical $b$ phases appearing with increasing pressure. This means that the necessary interactions for reproducing the observed spin structures are  contained in this {\it ab initio} model, and we can use this as a starting point for understanding the exchange interactions that stabilize the various phases as follows.

At $0$ GPa, the ferromagnetic state is stabilized at $J_\mathrm{min}=0.01$ meV. With increasing $J_\mathrm{min}$, the stabilized magnetic state changes to the helical $c$ state above $J_\mathrm{min}=1.0$ meV. The $J_\mathrm{min}$ dependence of the magnetic structure is similar up to pressures of about $3.20$ GPa. However, above $5.29$ GPa, the helical $b$ state is gradually stabilized. 
We can see that with increasing $J_\mathrm{min}$, the region where helical $c$ state is stabilized becomes larger.  This indicates that the helical $c$ state is stabilized by short-range magnetic interactions because increasing $J_\mathrm{min}$ corresponds to neglect of small $J$, i.e., long-range interactions. On the other hand, long-range magnetic interactions tend to stabilize the helical $b$ state.
Note that at $10.29$ GPa with $J_\mathrm{min} = 1.0$ meV, ferromagnetic state is stabilized. This is due to neglect of antiferromagnetic $J_3$, $J_4$ and $J_5$ with $J < 1.0$ meV as seen from Fig. ~\ref{fig:theor_J}. The $J_\mathrm{min}$ value that best reproduces the experimental phase diagram shown in Fig.~\ref{fig:phasediagram} (a) lies between $0.5$ and $1$ meV. Therefore, in the following, we set $J_\mathrm{min} = 0.5$ meV and investigate the temperature dependence of the stability of the magnetic states.

Figure \ref{fig:theor_phase}\correction{(a)} shows the temperature- and pressure-dependence of spin structures at $J_\mathrm{min} =  0.5$ meV. Here, we show the spin structures obtained by the snapshot at $1.8\times 10^5$ MC step. At $P=0$ GPa, the ferromagnetic state gradually changes to the paramagnetic state as the temperature is increased toward $300$~K. From $P=1.25$ to $3.20$ GPa, the period of helical $c$ state gradually becomes longer with increasing temperature and the magnetic state becomes ferromagnetic-like with a finite total magnetic moment around $200$~K and finally paramagnetic above $300$~K. The stabilisation of the ferromagnetic state near 200 K is  due to the fact that the antiferromagnetic interactions $J_4$ and $J_5$ with small magnitudes become negligible due to increased fluctuations at higher temperatures and $J_1$, $J_2$ and $J_3$ become dominant.
At $5.29$ GPa, the ferromagnetic state is stabilized again and becomes paramagnetic with increasing temperature toward $300$~K. This is expected to be due to a competition between the helical $b$ and $c$ states which cancel each other out and the remaining ferromagnetic interaction \correction{becomes}
dominant. Above $6.43$ GPa, the helical $b$ state is stabilized at $T=10$ K. The transition temperature from the helical $b$ state to the paramagnetic state also gradually decreases as a result of decreasing $J$ magnitudes with pressure.  \correction{We also checked the helical $c$ state for $1.25$ GPa and the helical $b$ state for 6.43 GPa at $10$ K are stabilized and the period lengths of each state do not change within a (20 nm)$^3$ cubic box.} 
In this way, the overall feature of the phase diagram in Fig. \ref{fig:phasediagram}, including the presence of each magnetic phase and the ferromagnetic and helical $b$ transition temperatures, is reproduced by the theoretical calculations, as shown in Fig. \ref{fig:theor_phase}\correction{(a)}. However, the critical pressures and the helical $c$ transition temperatures are not reproduced accurately. This may be partly due to Dzyaloshinskii–Moriya interactions and single ion anisotropy, which were neglected in the calculations.

\correction{As described above, we found that the propagation vector direction changes from the $c$ axis to the $b$ axis with applied pressure primarily because the long-range exchange interactions stabilize the helical $b$ state. On the other hand, magnetic anisotropies, such as the Dzyaloshinskii–Moriya interaction, are considered to play an important role in the change of the easy plane from the $ab$ plane to the $ac$ plane. As reported in Ref.~\cite{Shen2016}, the propagation vector decreases at the pressure where the easy-plane changes from the $ab$ plane to the $ac$ plane in CrAs. This behavior is explained considering the Dzyaloshinskii–Moriya interaction. Further studies are required to clarify the magnetic anisotropies in MnP.}

\section{Discussion} \label{sec:discussion}
Our high pressure experiments revealed that the helical $b$ is the stable magnetic structure in a wide pressure range from $\sim$3 GPa to the boundary of the superconducting phase.
\correction{Though the critical pressures and the helical $c$ transition temperatures are not reproduced accurately as described in the previous section, the} phase diagram obtained from our theoretical study (Fig.~\ref{fig:theor_phase}\correction{(a)}) is qualitatively consistent with the experiments (Fig.~\ref{fig:phasediagram}). Our analysis reveals that the change of the spin structure from helical $c$ to helical $b$ is related to the drastic change of the ferromagnetic interaction $J_3$. As shown in Fig. \ref{interactions}, $J_3$ corresponds to the Mn--Mn bond parallel to the $b$ axis. As the pressure is increased, $J_3$ becomes smaller (Fig.~\ref{fig:theor_J}), i.e., the ferromagnetic order in the $b$ direction is less favored with increasing pressure. On the other hand, the antiferromagnetic $J_5$ interaction which lies in the $ab$ plane remains mostly unchanged. These facts combine to favor antiferromagnetic helical order in the $b$ direction with increasing pressure, leading to the observed helical $c$--helical $b$ transition at low temperature. It is noted that the $J_1$--$J_2$--$J_4$ model, which has been employed to explain the helical spin structure at low pressure in several previous works (see Sec.~IIB), does not account for $J_3$ which we have elucidated as the most important ingredient leading to the helical $c$--helical $b$ transition.

Next, we discuss the pressure dependence of the experimentally observed incommensurability $\delta$ in the helical $b$ state. In the helical $c$ phase below 1.1 GPa, $\delta$ \correction{slightly} decreases with increasing pressure as shown in Fig. \ref{fig:moment_delta_P_dep}(b), suggesting that the ferromagnetic contributions become more dominant. This is consistent with the fact that the ferromagnetic phase is the ground state in the pressure range between 1.1 and 1.7 GPa. In the helical $b$ phase, $\delta$ at 3 K increases almost linearly from 0.09 r.l.u. at 1.8 GPa to 0.26 r.l.u. at \correction{7.5} GPa. The temperature dependence of $\delta$ in Fig. \ref{fig:moment_delta_P_dep}(c) shows that $\delta$ gradually increases with decreasing temperature.
This is consistent with lattice contraction as the sample cools.
The increasing $\delta$ with applied pressure indicates that ferromagnetic interactions are dominant at lower pressures and become less dominant with increasing pressure. Applying pressure increases itinerancy and makes antiferromagnetic interactions more dominant.
These observations are consistent with our theoretical results; our {\it ab initio} model shows that the ferromagnetic $J_1$ and $J_2$ interactions are suppressed with pressure (Fig.~\ref{fig:theor_J}), inducing shorter period helicity due to the relatively stronger antiferromagnetic interactions (Fig.~\ref{fig:theor_phase}).
\correction{The period lengths calculated from the experimental values are about $4.9$ nm at $0$ GPa in the helical $c$ state, while $2.2$ nm at $3.8$ GPa and $1.3$ nm at $7$ GPa in the helical $b$ state, respectively. As shown in Figs. \ref{fig:theor_phase}(b)-(d), with a $2$-fold period ($5$ nm) in the $c$-axis direction for helical c state and a $4$-fold period ($2.5$ nm) in the $b$-axis direction for helical $b$ state at $6.43$ GPa. The period of helical $b$ tends to shorten and disappear as the pressure increases. Though the long-period structure is sensitive to the lattice size, the results obtained in the present calculations are consistent with experimental results.}

\correction{It is noted that the interaction parameters obtained here are quite different from some of the previous works based only on theory and also obtained from fits to experimental 
data. The calculated $J$ values at ambient pressure are in good agreement with a previous work using the KKR formalism \cite{Tran2021}, although they differ significantly 
from a total-energy fit to the $J_1$--$J_2$--$J_4$ model ($J_1 \sim 66$ meV,  
$J_2 \sim -14$ meV, $J_4 \sim 84$ meV) \cite{Xu2017}. 
The $J$ values are also very different from those obtained by fitting an analytical 
expression to experimental spin wave excitations ($J_1 \sim 0.37$ meV, $J_2 \sim -0.65$ 
meV, $J_4 \sim -0.06$ meV, 
$J_5 \sim 0.26$ meV, $J_6 \sim 0.64$ meV) \cite{Yano2013,Itoh2014}. The discrepancies are not only in the magnitude of the parameters, but also the sign. In both of those 
works, the fitting was done on a model based on interactions up to a finite near-neighbor cutoff, while we have shown that much longer-ranged interactions are important for reproducing the phase transition behavior. This suggests that fitting with short-range models will result in convolution of the long-range interactions onto short-range $J$'s, and the results will depend on the range of interactions considered in the fitted model. In this case, the current result and the KKR result for $J$ values should be more reliable since they do not rely on an {\it a priori} model with a finite distance cutoff. We also suspect that the experimental work was at least partially 
affected by limited experimental data with large error bars. 
Moreover, the fact that our Monte Carlo simulations succeeded in reproducing the helical $c$/ferromagnetic/helical $b$ transitions and semi-quantitatively reproduced the 
incommensurability in the two helical phases lends credibility to our derived parameters. In particular, the abovementioned works do not consider the $J_3$ 
parameter, which we found to be a necessary ingredient in reproducing two distinct 
helical phases.}

Finally, we discuss a few remaining issues.
An interesting feature observed experimentally is that the $b$ axis expands slightly both at the ferromagnetic and helical transition temperatures.
The $b$ axis expansion at the ferromagnetic transition temperature is reasonable since it enhances the ferromagnetic $J_1$ and $J_3$ at low pressures (Fig. \ref{fig:theor_J}). On the other hand, the $b$ axis expansion at the transition to the helical $b$ phase may be related to the enhancement of antiferromagnetic $J_4$ and $J_5$ at high pressures.
In addition, to elucidate which exchange interactions become dominant in the vicinity of the superconducting phase and may couple to the superconducting pairing mechanism, the consideration of the electron itinerancy and coupling to the exchange interactions should be important.
In CrAs, an inelastic neutron scattering study was performed in the chemically pressurized system CrAs$_{1-x}$P$_x$ \cite{Matsuda2018}. In the nominal CrAs$_{0.94}$P$_{0.06}$, which is just above the critical doping level and shows no bulk antiferromagnetic order, antiferromagnetic fluctuations, which correspond to antiferromagnetic correlations coupled by $J_2$, were observed, suggesting a possible coupling between the magnetic fluctuations and the superconductivity. An NMR study also reported multiple kinds of magnetic fluctuations in the superconducting phase of CrAs at 1.09 GPa~\cite{Matsushima2019}.
In MnP, there has been no direct evidence to prove the presence of magnetic fluctuations around the superconducting phase. However, the quantum critical behavior and the non-Fermi-liquid behavior around the critical pressure to the superconducting phase~\cite{Cheng2015} suggest the presence of quantum magnetic fluctuations.
Our results suggest that antiferromagnetic $J_3$ may be one of the key interactions to give rise to the magnetic fluctuations.
Whether the pairing mechanism in the two materials is common or not merits further investigations.
These issues are beyond the level of theory considered in this work and may be tackled in future studies.

\section{Summary}
The pressure dependence of the magnetic structure in MnP has been elucidated and the pressure-temperature phase diagram has been completed. The helical $b$ structure is found to be robust in a wide pressure range between $\sim$3 and \correction{7.5} GPa. $\delta$ of the magnetic propagation vector along the $b$ axis increases almost linearly with applied pressure, suggesting that, with increasing pressure, antiferromagnetic contributions become more dominant in the frustrated interactions along the $b$ axis.
This behavior was explained clearly by our theoretical analysis, which showed that further neighbor interactions become more influential at higher pressures. \correction{The relative importance of the long-range component of the magnetic interaction means that the coupling between electrons and magnetism may be weakened and become more itinerant in the high-pressure region.}
Furthermore, $J_3$, nearest-neighbor interaction along the $b$ axis, shows the largest pressure dependence and experiences a drastic change from being ferromagnetic to antiferromagnetic, as the $b$ axis is contracted significantly with applied pressure.
\correction{The long-range magnetic interactions and } $J_3$ may be important interactions to understand the superconducting mechanism in MnP.
These results clearly demonstrate the synergy of combined neutron diffraction and theoretical studies to understand the magnetic properties under high pressures in detail. This will open a new avenue for exploring magnetic properties under high pressures.
We hope these results will stimulate further investigations into the hierarchy of the exchange interactions in the vicinity of the superconducting phase, and the possible means by which they may couple to the superconducting pairing mechanism.

\section*{Acknowledgments}
\correction{We would like to thank Dr. S. Yano for stimulating discussions.} This research used resources at the High Flux Isotope Reactor, a DOE Office of Science User Facility operated by the Oak Ridge National Laboratory. This work was partially supported by MEXT, the Grant-in-Aid for Scientific Research Grant No. 19H00648. \correction{This work was supported by JSPS KAKENHI Grant Number 21H01041.} This research used resources of the Advanced Photon Source, a U.S. Department of Energy (DOE) Office of Science user facility at Argonne National Laboratory and is based on research supported by the U.S. DOE Office of Science-Basic Energy Sciences, under Contract No. DE-AC02-06CH11357. SH acknowledges support provided by funding from the William M. Fairbank Chair in Physics and from the Fritz London Fellowship at Duke University. JGC is supported by the National Natural Science Foundation of China (Grant Nos. 12025408 and 11921004). JQY acknowledges the support by the U.S. Department of Energy, Office of Science, Basic Energy Sciences, Materials Sciences and Engineering Division.

\end{document}